\documentclass{article}
\usepackage[utf8]{inputenc}
\usepackage{listings}
\usepackage{float}
\usepackage{natbib}
\usepackage{graphicx}
\usepackage{amssymb}
\usepackage{amsmath}
\usepackage{mathtools}
\usepackage{listings}
\usepackage{color}
\usepackage{url}
\definecolor{dkgreen}{rgb}{0,0.6,0}
\definecolor{gray}{rgb}{0.5,0.5,0.5}
\definecolor{mauve}{rgb}{0.58,0,0.82}

\title{Blockchain Cohomology}
\author{Wyatt Meldman-Floch}
\date{%
    Constellation Labs\\%
    \today}
\begin{document}
\maketitle

\begin{abstract}
We follow existing distributed systems frameworks employing methods from algebraic topology to formally define primitives of blockchain technology. We define the notion of cross chain liquidity, sharding and probability spaces between and within blockchain protocols. We incorporate recent advancements in synthetic homology to show that this topological framework can be implemented within a type system. We use recursion schemes to define kernels admitting smooth manifolds across protocol complexes, leading to the formal definition of a Poincare protocol.

\end{abstract}

\section{Consensus Protocols}
Recent advancements in distributed computing adopt methods from algebraic topology for formally defining protocols \footnote{T. Nowak, "Topology in Distributed Computing" University of Vienna, \url{https://pdfs.semanticscholar.org/fd74/ed78ccffc6faa708b933fb0bdf7ceb62896d.pdf}}\footnote{M. Herlihy et al. \url{http://www.lix.polytechnique.fr/~goubault/papers/sv.pdf}}. We use these methods to model blockchain protocols as well as an internet of blockchains. We first define an execution space as a topological space equipped with a discrete product topology\footnote{Alpern, Bowen and Fred B. Schneider. Defining liveness. Technical Report TR85-650,
Cornell University, 1985. \url{https://ecommons.cornell.edu/bitstream/handle/1813/6495/85-650.pdf?sequence=1&isAllowed=y}}. Defining a distributed process in terms of topology only requires us to care about the structure of the set of possible schedules of a distributed system\footnote{Saks, Michael and Fotios Zaharoglou 2000, impossibility of the wait free k-set agreement \url{http://courses.csail.mit.edu/6.852/08/papers/SaksZaharoglu.pdf}}. We adopt Nowak's algebraic definition of an execution space in terms of the homology of protocol complexes \footnote{M. Herlihy et al.}. We define a protocol complex $S_k: P_k{\Delta^q}$ as the q-dimensional standard simplex
\begin{equation} \label{eq1}
\Delta^q = \{x \in \mathbb{R} | \Sigma x_j = 1, x_j \geq \forall j \}
\end{equation} \label{eq1}
at morphism $k$ described by the following vertex set
\begin{equation} \label{eq1}
S_k = \{v_{i,0} \dots v_{i,q}\}
\end{equation} \label{eq1}
where $P \subset S$ is the set of all admissible configurations and $S$ is the set of all possible configurations.
 
We define a consensus protocol $P^\sigma_{*}(S):\{S_k, \partial_k\}$ as the singular homology of a simplicial chain complex, carried by a group morphism implementing distributed consensus. Let $S_k$ be a simplex configuration at step $k$ and $\partial_k$ be the differential of a distributed consensus morphism:
\begin{equation} \label{eq1}
P^\sigma_{*}(S): 0 \leftarrow \dots P^\sigma(S_{k-1})\xleftarrow {\partial_{k-1}} P^\sigma(S_{k})\xleftarrow {\partial_{k}} P^\sigma(S_{k+1}) \dots
\end{equation} \label{eq1}
where $P_k = ker \partial_k / im \partial_{k+1}$ and is also an abelian group. Thus, $P_*= (P_k) \ | k \in \mathbb{Z}$ is a graded abelian group which is referred to as the homology of a protocol complex $S$. We abuse our notation of $P$ but rectify by noting that an admissible state $k$ is required for anther step $k + 1$, thus we define $P$ as the functor carrying our consensus operator defined below.

Define a consensus operator $\sigma$ as the group morphism on the singular q-simplex $\sigma: \Delta^q \rightarrow S$ 
\begin{equation} \label{eq1}
\sigma_k: S_{k-1} \times P_k \rightarrow S_{k}
\end{equation} \label{eq1}
which are continuous on discrete topologies\footnote{Nowak, Lemma 4.5} such as $\Delta^q$. Define homology between configurations as a measure of divergence given by the differential 
\begin{equation} \label{eq1}
\partial_k(\sigma) = \sum^{q}_{k=0} (-i)^{i-1}(\sigma \circ \delta_q^{i} )
\end{equation} \label{eq1}
for continuous functions $\delta^{i}_q: \Delta^{q-1} \rightarrow \Delta^q | 1 \leq i \leq q+1$ where 
\begin{equation} \label{eq1}
\delta^{i}_q(x_1, \dots, x_q) = (x_1, \dots x_{i-1}, 0, x_i, x_{i+1}, \dots, x_{q-1}, \dots, x_q)
\end{equation} \label{eq1}

As the graded abelian group of our protocol complex is the simplicial singular homology group and $\sigma$ is our homology preserving map, it is trivial to note that homology holds $\forall k \in \mathbb{Z}$, i.e.
\begin{equation} \label{eq1}
\partial_k \circ \partial_{k+1} = 0
\end{equation} \label{eq1}

As a corollary of the fact that the geometric realization of a simplicial complex is dually a topological space, due to the vanishing cohomology up to $k$, we note that $P_k\Delta^q$ is k-acyclic\footnote{Nowak, Definition 5.4}.

\section{Protocol Topologies}
It's possible that we could 'mix' protocol complexes defined as above. We employ our notion of cohomology to define a 'liquidity' or the ability to exchange configuration states between protocol complexes. We leave applications of this as an exercise for the reader.

We define liquidity as the existence of a functoral vertex map between singular homologies (defined equivalently here as the disjoint subset of protocol complexes) $l: \bigcup_{k} P_{\pi} \rightarrow \bigcup_{k} P_{\pi+1}$.

Making use of homotopy type theory allows us to focus on structure by treating topological characteristics called homotopy groups as primitives. If we redefine our k-acyclic distributed consensus protocol $\sigma$ categorically as the functoral carrier $\Sigma_{*}$  we can form a chain complex that adheres to the homology theory of homotopy types\footnote{R. Grahm "Synthetic Homology in Homotopy Type Theory" \url{https://arxiv.org/pdf/1706.01540.pdf}}

Simplicial complexes together with simplicial vertex maps form a category. Let us define a protocol topology $T^{\Sigma}_P: $ $\Sigma_{*}P_\pi$ as the singular homology of a chain complex of protocol complexes carried by a homotopy preserving functor $\Sigma_*$. The protocol topology is given by the following chain complex

\begin{equation} \label{eq1}
T^{\Sigma}_P: 0 \leftarrow \Sigma_{*}P_\pi \xleftarrow{\partial} \Sigma P_{0} \xleftarrow{\partial} \dots \Sigma P_i \ | i \leq \pi \in \mathbb{Z}
\end{equation} \label{eq1}
where $\Sigma_\pi: $ $ker \partial^{\pi}_{k}/im \partial^{\pi}_{k+1} \rightarrow \partial^{\pi+1}_{k} /im \partial^{\pi+1}_{k+1}$ 

For protocol complex morphisms $\Sigma_\pi, \Sigma_{\pi+1}$ chain homotopy from $\Sigma_\pi$ to $\Sigma_{\pi+1}$ is a homotopy preserving graded abelian group morphism $l: P_{\pi} \rightarrow P_{\pi+1}$ yielding a vanishing homology, i.e. 

\begin{equation} \label{eq1}
\begin{split}
\Sigma_\pi - \Sigma_{\pi+1} =  \partial^{\pi}\circ l + l \circ \partial^{\pi+1} \\
= \partial^\pi \circ \partial^{\pi+1} = 0 \\
\end{split}
\end{equation}

Noting that these conditions are met by the definitions of an acyclic carrier \footnote{Nowak, Theorem 5.1}, it follows that a protocol topology as defined above is $\pi$-acyclic.

\section{Block Sheaves}
Designing distributed architectures with topology gives us a lot of power, but in order to use it we need to design our topologies such that they are mathematically tractable for solving a specific problem. In principle, Abstract Differential Geomoetry (ADT) admits any topological space as a base space on which to 'solder sheaves' for carrying out differential geometry\footnote{A. Mallios et al. "Finitary Cech-de Rham Cohomology: much ado without $C^{\infty}$-smoothness"}. We introduce methods from Abstract Differential Geometry, namely finitary cech-deRham cohomology in order to define an orientable manifold from our definition of protocol topology.

First we need to introduce the dual of homology as described above, namely cohomology. In describing our protocol complex it only makes sense to have an arrow moving 'forward in time' as consensus itself is acyclic, with each iteration pointing 'backwards in time' to its previous state. In this sense our evolution was the compounding dimensionality of the space of all configurations, as implied by the discrete product topology of a protocol complex. In defining an orientable manifold, we need to move 'backwards' through our  space, i.e. from higher to lower dimension. This is shown as the differential on an arrow going right instead of left.

By constructing the protocol topology within a monoidal category, the singular cohomology of a protocol topology is equivalent to an A-module of Z+-graded discrete differential forms. One can, in a natural way, assign a decision tree to any set of executions that captures the decision of choosing a successor\footnote{Nowak, Section 4.1.2}. A blockchain can be defined as an extension of an execution tree, where each block is formulated as a sheaf with a well defined tensor operation. We define a sheaf $\epsilon$ as the 'enrichment' of any cochain $\mathbb{A}$-complex of positive degree/grade, corresponding to the $\mathbb{A}$-resolution of an abstract $\mathbb{A}$-module
\begin{equation} \label{eq1}
S^*: 0 \rightarrow \epsilon \rightarrow S^0 \xrightarrow{d^0} S^1 \xrightarrow{d^1} \dots
\end{equation} \label{eq1}
and homomorphism given by  Cartan-Kahler-type of nilpotent differential operator d. We will make use of the fact that an $\mathbb{A}$-module sheaf $\epsilon$ on any arbitrary topological space (shown above with an arbitrary simplicial cochain-complex) admits an injective resolution per (10).

Blockchains are naturally equipped with a sheaf, that of the block.  This would allow us to 'unpack' data within a block recursively under the product operation. Every abelian unital ring admits a derivation map \footnote{Mallios, A., Geometry of Vector Sheaves: An Axiomatic Approach to Differential
Geometry, vols. 1-2, Kluwer Academic Publishers, Dordrecht (1998)}, thus if we reformulate our definition of a consensus protocol above as a sheaf with semigroup operations carried by right derived functors with monadic bind, we can form a manifold.

By noting the equivalence of Sorkin's fintoposets\footnote{Section 3.2, Mallios et al.} as simplicial complexes, Mallios et al. showed that the Gelfand duality\footnote{Section 3.3, Mallios et al.} implies that a manifold can be constructed out of the incidence Rota algebra of a simplex's corresponding fintoposet \footnote{eq 9, Mallios et al.}. For a fintoposet (the topological equivalent of a directed acyclyc graph), it's incidence algebra can be broken down into a direct sum of vector subspaces
\begin{equation} \label{eq1}
\Omega(P) = \bigoplus_{i \in \mathbb{Z}_+} \Omega^i = \Omega^0 \oplus \Omega^0 \dots := A \oplus R
\end{equation} \label{eq1}
where $\Omega(P)$s are $\mathbb{Z}_+$ graded linear spaces, $A$ is a commutative sub algebra of $\Omega$ and $R := \bigoplus_{i \geq 1} \Omega^i$ is a linear (ringed) subspace. It is trivial to notice that $\Omega(P)$ is an A-module of a Z+-graded discrete differential form. 

A manifold can be constructed by organizing the incidence algebras of our protocol complexes into algebra sheaves. The n-th (singular) cohomolgy group $H_n(X, \epsilon)$ of an A-module sheaf $\epsilon(X)$ over topological space $X$, can be described by global sections $\Gamma_X(\epsilon) \equiv \Gamma (X, \epsilon)$
\begin{equation} \label{eq1}
H_n(X, \epsilon) := R^n(\Gamma(C,\epsilon) := H^n[\Gamma(C, S^*)] := ker\Gamma_X(d^n)/im\Gamma_X(d^{n-1})
\end{equation} \label{eq1}
where $R^n\Gamma$ is the right derived functor of the global section functor $\Gamma_x(.) \equiv \Gamma(X,.)$. Note that $R^n$ is equivalent to the $i^{th}$ linear ringed subspace above. These dual definitions of gamma correspond to out definitions of $\sigma$ and $\Sigma_*$ with respect to our functoral vertex map $l$ in our definition of a protocol topology.

The sheaf cohomology of a topological space is the cohomology of any $\Gamma_X$-acyclic resolution of $\epsilon$\footnote{Mallios, A., "On an Axiomatic Treatment of Differential Geometry via Vector
Sheaves." Applications, Mathematica Japonica }. The corresponding abstract $\mathbb{A}$-complex $S^*$ can be directly translated by the functor $\Gamma_x$ to the 'global section $\mathbb{A}$-complex' $\Gamma_X(S^*)$
\begin{equation} \label{eq1}
\Gamma_X(S^*):  0 \xrightarrow{~} \Gamma_X(\epsilon) \xrightarrow{d^0} \Gamma_X(S^0) \xrightarrow{d^1} \dots
\end{equation} \label{eq1}
which is the abstract de Rham complex of a discrete manifold $X$. The action of d is to effect transitions between the linear subspaces $\Omega_i$ of $\Omega(P)$ in (11), as follows: d: $\Omega_i \rightarrow \Omega_{i+1}$.

The finitary de Rham theorem defines a finitary equivalent of the typical $c^{\infty}$ smooth manifold. Noting $\Gamma^{P_m}_m$ is fine by construction, Mallios et al. show that finsheaf-cohomology differential tetrads
\begin{equation} \label{eq1}
\tau := (P_m, \Omega_M, d, \Omega^M_{deR})
\end{equation} \label{eq1}
is equivalent to the $c^{\infty}$-smooth Cech-de Rham complex. In our definition of $\tau$, $\Omega_M$ is the categorically dual finsheaf (finitary sheaf) of Sorkin's  fintoposets $P_m$, d is effectively an exterior product, and $ \Omega^M_{deR}$ is the abstract de Rahm complex.

\section{Blockchain Cohomology}
We've shown how to create a manifold from the cohomology of a discrete topological space.  We can define a synthetic manifold out of a protocol topology\footnote{J. Gallier et al. Definition 3.3: "A Gentle Introduction to Homology, Cohomology, and
Sheaf Cohomology" https://www.seas.upenn.edu/~jean/sheaves-cohomology.pdf}. Define a cochain-complex within the cohomology theory of homotopy types under the cup product.

Making note of the existence of a tensor product in the cohomology theory of homotopy types by E. Cavallo \footnote{E. Cavallo, "Synthetic Cohomology in Homotopy Type Theory", \url{http://www.cs.cmu.edu/~ecavallo/works/thesis15.pdf}} we define the protocol manifold as 
\begin{equation} \label{eq1}
\Gamma^\epsilon_{\Sigma} = \bigoplus_{0 \leq i \leq \pi} \Sigma_* \epsilon_i
\end{equation} \label{eq1}

\section{Typesafe Poincare Duality}
Up until now we have not explicitly defined functoral group homomorphisms that can construct the complexes described above. We show that the dual nature of the hylomorphic and metamorphic recursion schemes maintain vanishing differentials and thus poincare duality for all $\pi$.

If we define a catamorphism and anamorphism with the same f-algebra and f-coalgebra, we can show by construction that the resulting co/chain-complexes are valid definitions of protocol topologies/manifolds and that poincarre duality of the protocol manifold is maintained up to $\pi$ isomorphism. We define in terms of $\Sigma$ and $\epsilon$, noting that our functor $\Sigma$ is a valid f-algebra and sheaf $\epsilon$ a co-algebra.

Let us define a hylomorphism
\begin{equation} \label{eq1}
\epsilon \leftarrow P \times \Sigma  : \Omega^T(\epsilon, P)
\end{equation} \label{eq1}
and metamorphism
\begin{equation} \label{eq1}
\Omega_\Gamma(P, \epsilon):\Gamma_\Sigma \times \epsilon \rightarrow P  
\end{equation} \label{eq1}
we formally verify by the construction of the following geometric cw-complex
\begin{equation} \label{eq1}
\Omega^T_\Gamma: 0 \xleftrightarrow{\partial} \Omega^{T^*}_{\Gamma^*}(\epsilon) \xleftrightarrow{\partial}  \Omega^T_\Gamma (\epsilon(P_0)) \dots \Omega^T_\Gamma (\epsilon(P_\pi))
\end{equation} \label{eq1}
that $T$ and $\Gamma$ form a poincare complex, clearly satisfying the poincare duality as $\partial$ vanishes in our construction of $T$ and $\Gamma^\epsilon_\Sigma$. The fundamental class of our corresponding space is $\Omega^{T^*}_{\Gamma^*}$ which carries the type signatures of our hylo and metamorphisms. Formally define $\Omega^{T}_{\Gamma}$  as a Poincare protocol.

\section{Remarks} 
It's worth noting that the isomorphism between simplectic and poset topology shown by Sorkin's fintoposets implies that when applied to blockchains, the existence of cycles in a cohomological or homological cw-complex imples the existence of forks. In future work we will show that finite autamata with monoidal state transitions (semiautomation) admit a Poincare protocol with enrichment isomorphic to the semigroup operation of state transitions. Extending this, we'll make use of a Poincare protocol's manifold to define monoidal state transitions that prevent divergences, and transitively forks.

\bibliographystyle{plain}
\end{document}